\documentclass[11pt,twoside]{article}
\usepackage{FUSE2004}
\usepackage{psfig}
\usepackage{epsf}
\usepackage{graphics}
\usepackage{graphicx}
\usepackage{lscape}
\usepackage{wrapfig}
\def\edcomment#1{\iffalse\marginpar{\raggedright\sl#1\/}\else\relax\fi}
\reversemarginpar

\newcommand{\msolar}{\mbox{\,$M_{\odot}$}}

\begin{document}
\title{\ion{O}{6} Column Density Distribution in the Local Bubble - Results from 3D Adaptive Mesh Refinement Simulations}

\author{Miguel A. de Avillez}
\affil{Department of Mathematics, University of \'Evora, R. Rom\~ao
Ramalho 59, 7000 \'Evora, Portugal. {\it
Email:}mavillez@galaxy.lca.uevora.pt}
\author{Dieter Breitschwerdt}
\affil{Institut f\"ur Astronomie, Universit\"at Wien, T\"urkenschanzstr. 17, A-1180 Wien, Austria. {\it Email:} breitschwerdt@astro.univie.ac.at}

\begin{abstract}
The Local Bubble (LB) is an X-ray emitting region extending 100 pc in
radius in the Galactic plane and 300 pc perpendicular to it, and it is
embedded in a somewhat larger H{\sc i} deficient cavity. Its origin
and spectral properties in UV, EUV and X-rays are still poorly
understood.  We have performed 3D high resolution (down to 1.25 pc)
hydrodynamic superbubble simulations of the LB and Loop I superbubble
in a realistic \emph{inhomogeneous} background ISM, disturbed by
supernova (SN) explosions at the Galactic rate. We can reproduce (i)
the size of the bubbles (in contrast to similarity solutions), (ii)
the interaction shell with Loop I, discovered with ROSAT, (iii)
predict the merging of the two bubbles in about 3 Myr, when the
interaction shell starts to fragment, and, (iv) the generation of
blobs like the Local Cloud as a consequence of a dynamical
instability. The \ion{O}{6} column densities are monitored and found
to be in excellent agreement with \textsc{Copernicus} and FUSE
absorption line data, showing LB column densities $<1.7 \times 10^{13}
\, {\rm cm}^{-2}$, in contrast to other existing models.
\end{abstract}
\thispagestyle{plain}

\section{Introduction}
Standard Local Bubble (LB) models fail to reproduce the observed low \ion{O}{6}
 absorption column density (Shelton \& Cox 1994; for a recent discussion
see Breitschwerdt \& Cox 2004). Heliospheric in situ measurements are
sensitive to the boundary conditions imposed by the Local Bubble and
the \ion{O}{6} column density in absorption is a crucial test for modelling
of the local ISM (Cox 2004).  It seems most plausible that the LB is
theresult of 20 successive explosions, originating from stars with
masses between 11 and 20 $\msolar$ from the moving subgroup B1 of the
Pleiades in the last 14.5 Myrs (Bergh\"ofer \& Breitschwerdt 2002). We
have simulated the LB evolution by means of a 3D AMR hydrocode and
calculated the observed \ion{O}{6} column density in absorption
(along lines of sight crossing Loop I).

\section{Model and Simulations}

We use the 3D model of Avillez (2000), where the ISM is disturbed by
supernova (SN) explosions at the Galactic rate, and took data cubes of
previous runs with a finest adaptive mesh refinement resolution of 1.25
pc (Avillez \& Breitschwerdt 2004). We then picked up a site with enough
mass to form the 81 stars, with masses, $M_*$, between 7 and 31
$\msolar$, that compose the Sco Cen cluster; 39 massive stars with $14
\leq M_* \leq 31 \, \msolar$ have already gone off, generating the Loop
I cavity.
\begin{wrapfigure}[16]{l}[0pt]{6.cm}
\centering

\vspace*{2cm}
Jpeg image 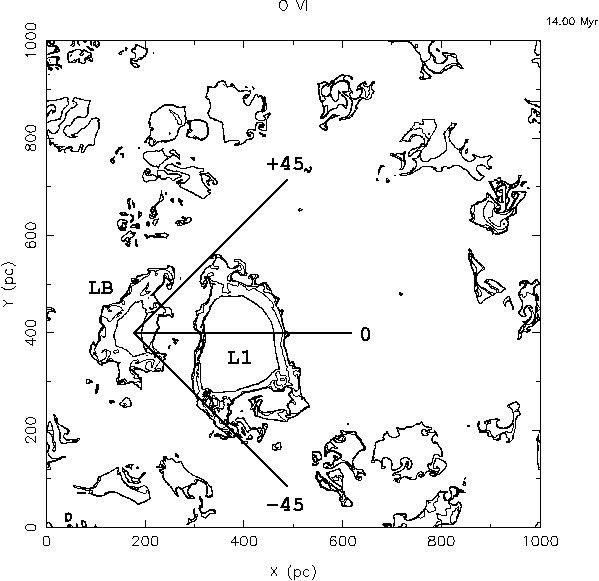
 \vspace*{2.6cm}

\hspace*{-0.6cm}\begin{minipage}[l]{7cm}

\caption{\ion{O}{6} contour map of a 3D Local Bubble simulation 14 Myr after the first
    explosion; LB is centered at (175, 400) pc and Loop I
    at (375, 400) pc.}
\end{minipage}
\label{fig1}
\end{wrapfigure}
Presently the Sco Cen cluster (here located at $(375,400)$ pc) has 42
stars to explode in the next 13 Myrs). We followed the trajectory of
the moving subgroup B1 of Pleaides, whose SNe in the LB went off along
a path crossing the solar neighbourhood (Fig.~1). Periodic boundary
conditions are applied along the four vertical boundary faces, while
outflow boundary conditions are imposed at the top ($z=10$ kpc) and
bottom ($z=-10$ kpc) boundaries. The simulation time of this run was
30 Myr.

\section{Results}
\textbf{Morphology:} The locally enhanced SN rates produce coherent
LB and Loop I structures (due to ongoing star formation) within a
highly disturbed background medium. The successive explosions heat
and pressurize the LB, which at first looks smooth, but develops
internal temperature and density structure at later stages. After 14
Myr the LB cavity, bounded by an outer shell, which will start to
fragment due to Rayleigh-Taylor instabilities in $\sim 3$ Myr from
now, fills a volume roughly corresponding to the present day size
(Fig.~1).
\begin{figure}[bh]
\centering
\includegraphics[width=0.4\hsize,angle=0]{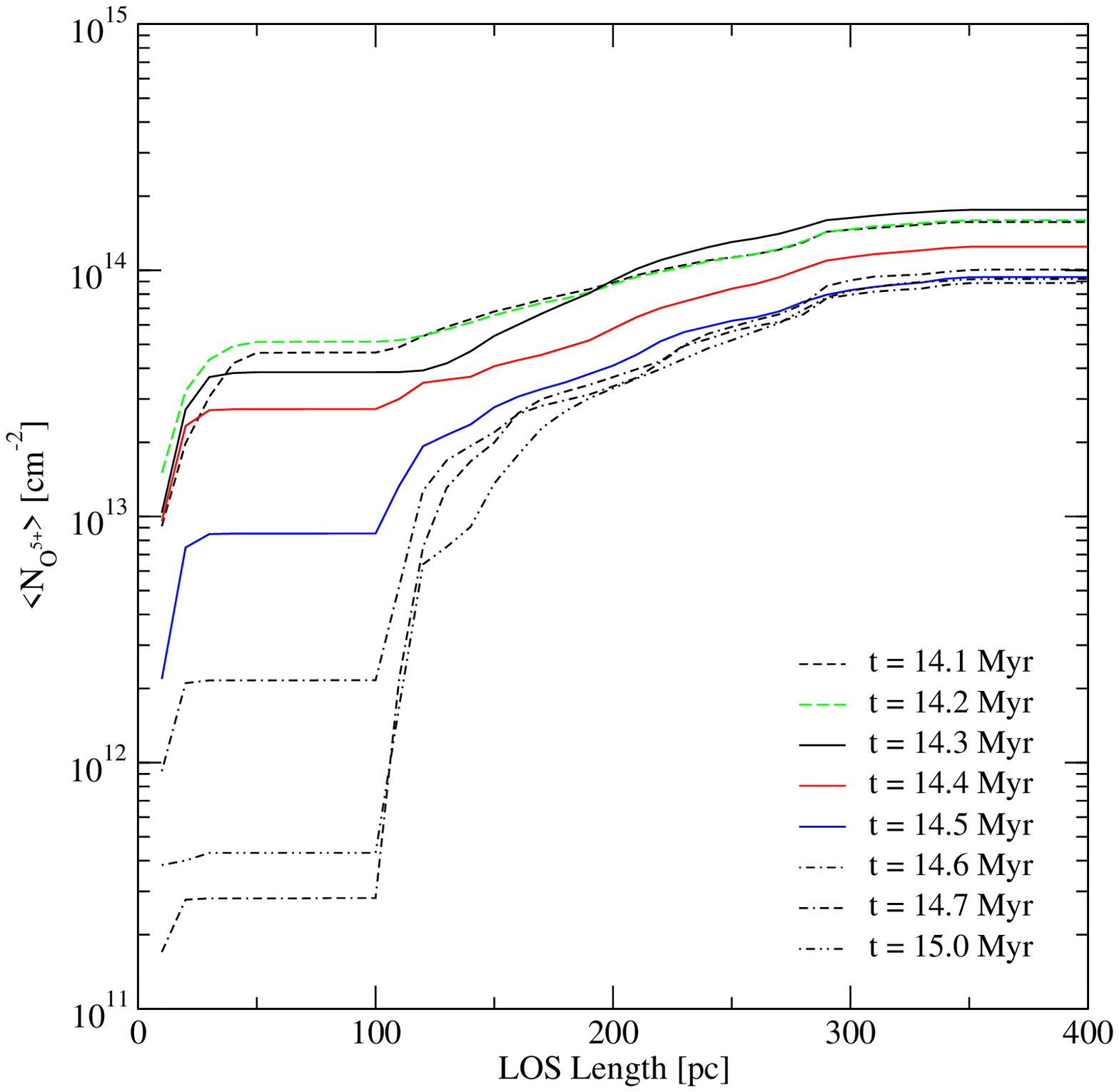}
\includegraphics[width=0.4\hsize,angle=0]{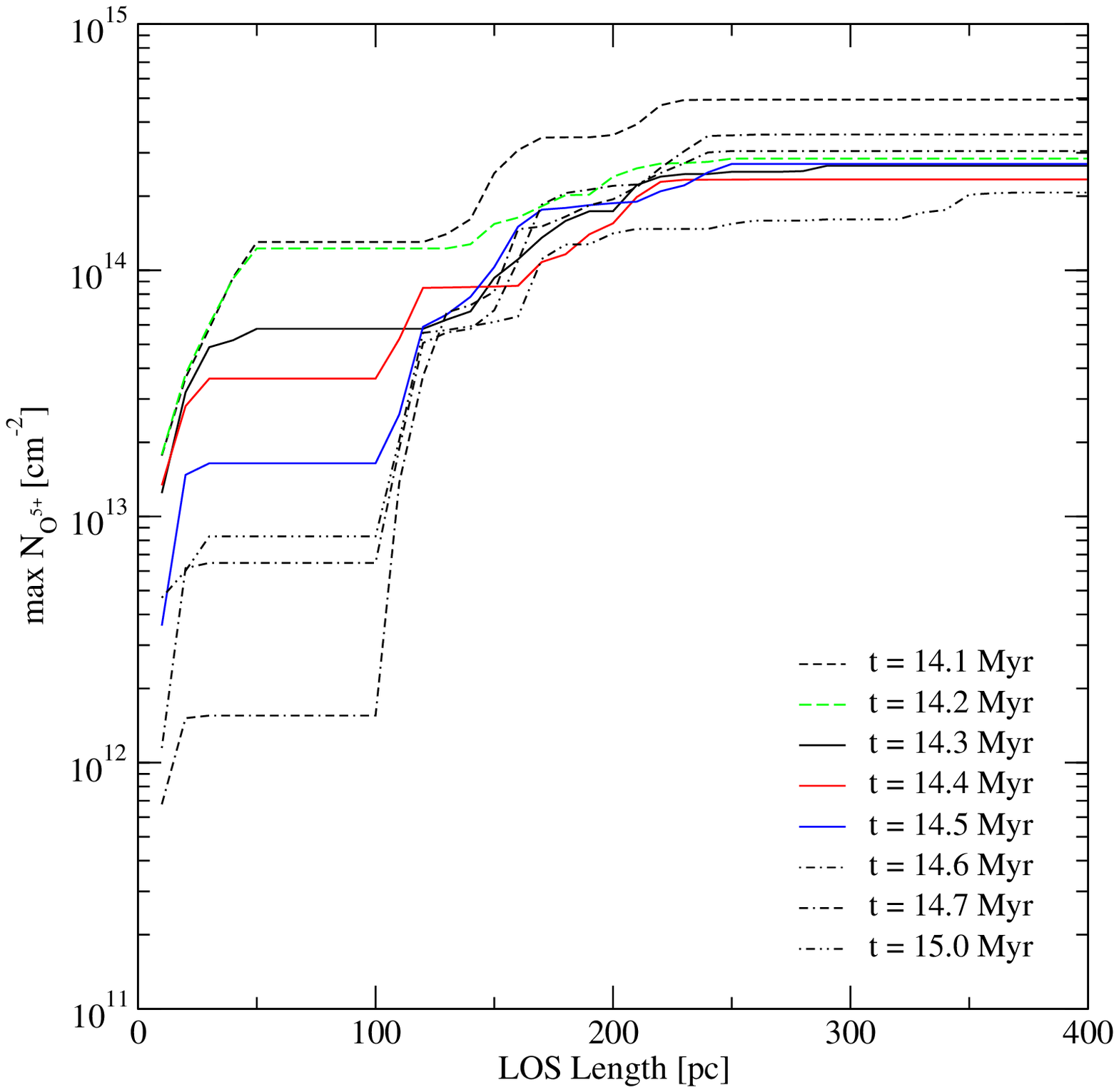}
\caption{
\ion{O}{6} column density averaged over angles (left panel) indicated
in Fig.~\ref{fig1} and maximum column density (right panel) as a
function of LOS path length at $14.1 \leq t\leq 15$ Myr of Local and
Loop I bubbles evolution.}
\label{fig3}
\end{figure}

\textbf{Oxygen {\sc vi}:} The average and maximum column densities of
\ion{O}{6}, i.e., $\langle \mbox{N(\ion{O}{6})} \rangle$ and
$\mbox{N}_{\mbox{max}}\mbox{(\ion{O}{6})} $ are calculated along 91
lines of sight (LOS) extending from the Sun and crossing Loop I from
an angle of $-45\deg$ to $+45\deg$ (s.~Fig.~1). Within the LB (i.e.,
for a LOS length $l_{LOS}\leq 100$ pc) $\langle \mbox{N(\ion{O}{6})}
\rangle$ and $\mbox{N}_{\mbox{max}}\mbox{(\ion{O}{6})} $ decrease
steeply from $5\times10^{13}$ to $3\times 10^{11}$ cm$^{-2}$ and from
$1.2\times 10^{14}$ to $1.5\times 10^{12}$ cm$^{-2}$, respectively,
for $14.1 \leq t\leq 15$ Myr (Fig.~2), because no further SN
explosions occur and recombination is taking place. For LOS sampling
gas from outside the LB (i.e., $l_{LOS}>100$ pc) $\langle
\mbox{N(\ion{O}{6})} \rangle > 6\times 10^{12}$ and
$\mbox{N}_{\mbox{max}}\mbox{(\ion{O}{6})} > 5\times 10^{13}$
cm$^{-2}$. The histograms of column densities obtained in the 91 LOS
for $t=14.5$ and 14.6 Myr (Fig.~3) show that for $t=14.6$ Myr all the
LOS have column densities smaller than $10^{12.9}$ cm$^{-2}$, while
for $t=14.5$ Myr 67\% of the lines have column densities smaller than
$10^{13}$ cm$^{-2}$ and in particular 49\% of the lines have
$\mbox{N(\ion{O}{6})}\leq 7.9\times 10^{12}$ cm$^{-2}$.
\begin{wrapfigure}[16]{l}[0pt]{6.0cm}
\centering
\centerline{\includegraphics[width=0.9\hsize,angle=0]{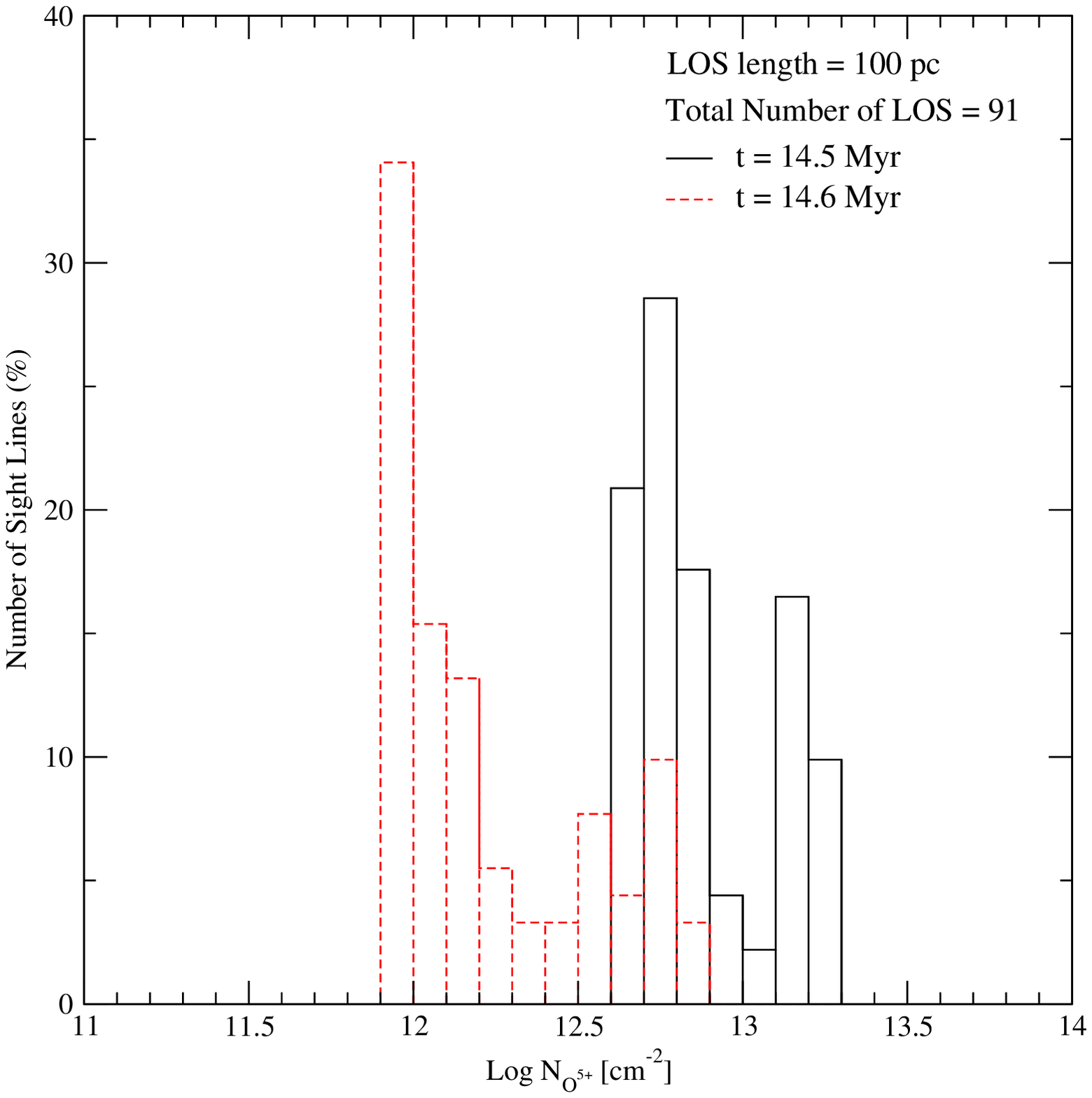}}
\hspace*{-0.6cm}\begin{minipage}[l]{7cm}
\caption{
Histogram of the percentage of LOS with various ranges of observed
$\mbox{N(\ion{O}{6})}$ within the LB at $t=14.5$ and $14.6$ Myr.}
\end{minipage}
\label{fig2}
\end{wrapfigure}
Noting that in the present model for $t\geq 14.5$ Myr the \ion{O}{6} column
densities are smaller than $1.7\times 10^{13}$ cm$^{-2}$ and $\langle
\mbox{N(\ion{O}{6})} \rangle \leq 8.5\times 10^{12}$ cm$^{-2}$ (blue line
in both panels of Fig.~2) and that a mean column density of $7\times
10^{12}$ cm$^{-2}$ is inferred from analysis of \textsc{FUSE} absorption
line data in the Local ISM (Oegerle et al. 2004) we estimate that the
present LB time is $14.7_{-0.2}^{+0.5}$ Myr.

\section{Conclusions}

The \ion{O}{6} column density is a sensitive tracer of the age of an
\textit{evolved} superbubble, and can thus give a constraint on the
timescale of the last explosion, which occurred in the present
simulations $0.6_{-0.2}^{+0.5}$ Myr ago, whereas the nearest
explosions to the Sun happened $2.8\pm 0.7$ Myr ago, in good
agreement with the most recent dating inferred from $^{60}\mbox{Fe}$
isotope analysis in the ferromanganese crust of deep ocean layers
(Knie et al. 2004).

{\small {\bf Acknowledgements} M.A. would like to thank the
organization for the financial support to attend this excellent
conference.}


\begin{references}

\reference{}Avillez, M. A.\ 2000, MNRAS, 315, 479.
\reference{}Avillez, M. A., \& Breitschwerdt,~D.\ 2004, A\&A, 425, 899
\reference{}Bergh\"ofer, T., \& Breitschwerdt, D.\ 2002, A\&A, 390,
299.
\reference{}Breitschwerdt, D., \& Cox, D. P.\ 2004, in ``How does the
Galaxy Work?'', eds. E. Alfaro, E. Perez, \& J. Franco, Kluwer
(Dordrecht), p. 391
\reference{}Cox, D. P.\ 2004, Ap\&SS, 289, 469
\reference{}Knie, K., et al.\ 2004, Phys. Rev. Lett., 93, 17
\reference{}Oegerle, W. R., et al. \ 2004, ApJ, submitted [astro-ph/0411065]
\reference{}Shelton, R., \& Cox, D. P.\ 1994, ApJ, 434, 599
\end{references}
\end{document}